\documentclass[11pt]{article}

\usepackage{srcltx}
\usepackage[dvips,letterpaper]{geometry}
\usepackage{graphicx,psfrag,epsf}
\usepackage{bbm}
\usepackage{times}
\usepackage{fullpage}
\usepackage{setspace}
\usepackage{amsmath,amssymb,amsfonts}
\usepackage{bm}
\usepackage{graphicx}
\usepackage{epsfig}
\usepackage{subfigure}
\usepackage{url}
\usepackage{verbatim}
\usepackage{natbib}
\usepackage{color}
\usepackage{booktabs}
\usepackage{graphicx,psfrag,epsf}
\usepackage{longtable}

\usepackage{physics}
\usepackage{lineno}
\usepackage{here}
\usepackage{lscape}
\usepackage{booktabs}
\usepackage{amsfonts}

\input cyracc.def

\usepackage{amsthm}
\newtheorem{example}{Example}

\newtheorem{Remark}{Remark}[section]

\title{Fractional approaches for the distribution of innovation sequence of INAR(1) processes}

 \vspace{-0.25cm}
\author{\normalsize
\textbf{
Josemar Rodrigues$^{1}$\thanks{Corresponding
author: Josemar Rodrigues. Email: josemar@icmc.usp.br }, \,
Marcelo Bourguignon}$^{2}$,\, \textbf{Manoel Santos-Neto}$^{3, 4}$\, and \, \textbf{N. Balakrishnan}$^{5}$
\\
{\footnotesize $^{1}$Departamento de Estat\'istica, Universidade de S\~ao Paulo, S\~ao Carlos, Brazil}\\[-0.15cm]
{\footnotesize $^{2}$Departamento de Estat\'istica, Universidade Federal do Rio Grande do Norte, Natal, Brazil}\\[-0.15cm]
{\footnotesize $^{3}$Departamento de Estat\'istica, Universidade Federal de Campina Grande, Campina Grande, Brasil}\\[-0.15cm]
{\footnotesize $^{4}$Departamento de Estat\'istica, Universidade Federal de S\~ao Carlos, S\~ao Carlos, Brasil}\\[-0.15cm]
{\footnotesize $^{5}$Department of Mathematics and Statistics, McMaster University, Ontario, Canada L8S4KI}\\[-0.15cm]}

\date{}

\begin{document}

\maketitle

\vspace{-0.7cm}
\begin{abstract}
In this paper, we present a fractional decomposition of the probability generating function of the innovation process
of the first-order non-negative integer-valued autoregressive [INAR(1)] process to obtain the corresponding probability mass function.
We also provide a comprehensive review of integer-valued time series models, based on the concept of thinning operators,
with geometric-type marginals.
In particular, we develop four fractional approaches to obtain the distribution of innovation processes of the INAR(1) model
and show that the distribution of the innovations sequence has geometric-type distribution. These approaches are discussed in detail and illustrated through a few examples. Finally, using the methods presented here, we develop four new first-order non-negative
integer-valued autoregressive process for autocorrelated counts with overdispersion with known marginals, and derive some  properties of
these models.

\vspace{-0.3cm}

\end{abstract}

\vspace{-0.3cm}
\section{Introduction}\label{sec:1}

Models for count time series with overdispersion, based on thinning
operators, have been discussed by many authors.
The vast majority of the proposed integer-valued autoregressive processes are based on geometric-type distribution.
There is an exhaustive list of INAR(1) processes with geometric-type marginal
distributions, geometric-type count variables or geometric-type innovations.

\cite{McKenzie:1986aa} and \cite{Al-Osh:1992} proposed  INAR(1) processes with geometric and negative binomial distributions as marginals.
Similarly, \cite{Alzaid:1988} discussed the geometric INAR(1) process [GINAR(1)]. \cite{Ristic:2009aa} introduced the geometric
first-order integer-valued autoregressive [NGINAR(1)] model with geometric marginal distribution.
Integer-valued time series, with geometric marginal distribution, generated by mixtures of binomial and negative binomial thinning
operators have been considered by \cite{Nastic:2012aa}, \cite{Nastic:2012ab} and \cite{Ristic:2012aa}. Recently, \cite{Nastic:2016ac} constructed
a new stationary time series model with geometric marginals, based on thinning operator, which is a mixture of Bernoulli
and geometric distributed random variables.

Of course, a simple approach is to change the distribution of innovations. In this context, \cite{Jazi:2012aa} proposed
the geometric INAR(1) model with geometric innovations, while \cite{Bourguignon:2018} introduced a first-order non-negative integer-valued
autoregressive process with zero-modified geometric innovations based on binomial thinning.

In context of dependent Bernoulli counting variables, \cite{Ristic:2013aa}, \cite{Ilic:2016aa} and \cite{Nastic:2017aa}
introduced an integer-valued time series model with geometric marginals based on dependent Bernoulli count variables.
Recently, \cite{MileticIlic:2017aa} introduced an INAR(1) model based on a mixed dependent and independent count series
with geometric marginals.

\cite{Nastic:2016aa} introduced an $r$-states random environment non-stationary INAR(1)
which, by its different values, represents the marginals selection mechanism from a family of different geometric distributions.
\cite{Borges:2016aa} and \cite{Borges:2017aa} introduced  geometric first-order integer-valued autoregressive processes with geometric
marginal distribution based on $\rho$-binomial thinning operator and $\rho$-geometric thinning operator, respectively.

Other works that have recently appeared in the literature dealing with geometric-type INAR(1) processes are those
of \cite{Nastic:2012ac} (shifted geometric INAR(1) process), \cite{Barreto-Souza:2015ab} (INAR(1) process with zero-modified geometric marginals) and \cite{Yang:2016aa} (threshold INAR(1) process).
Final mention should be made to the work of \cite{Ristic:2012ab}, \cite{Popovic:2016ab} and \cite{Popovic:2016aa} proposed bivariate INAR(1) time series models with geometric marginals.

In many recent works, it is not clear how the distribution of the innovation processes was derived.
In this context, motivated by \citet[see][p. 276]{feller08aa}, we formulate here some procedures to obtain the probability
mass function of the innovation process of an INAR(1) process by rewriting the probability generating function (pgf)
of the innovation process as a quotient of two polynomial functions of enequal degrees. In particular, we present four fractional
approaches to obtain the probability mass function of innovation processes of the INAR(1) model when the distribution of the
innovations sequence has geometric-type distribution. Furthermore, we put forward four new first-order non-negative integer-valued
autoregressive processes (Examples 6, 7, 8 and 9) with inflated-parameter Bernoulli and inflated-parameter geometric marginals \citep[][]{kolev2000}.

The rest of this paper is organized as follows. In Section 2, we develop four fractional approaches to obtain the distribution of innovation processes of the INAR(1) model and show that the distribution of the innovations sequence has geometric-type distribution. In addition, some illustrative
examples and new  models are presented. Finally, some concluding remarks are made in Section 3.

\section{Main results}
\label{sec:2}

Let $\mathbb{Z}$, $\mathbb{N}$ and $\mathbb{R}$ denote the set of integers,
positive integers and real numbers, respectively. All random variables will be
defined on a common probability space \ $(\Omega,\mathcal{A},\mathbb{P})$. A discrete-time stationary non-negative integer-valued stochastic process
$\{X_{t}\}_{t\in \mathbb{Z}}$ is said to be a first-order integer-valued autoregressive [INAR(1)]
process if it satisfies the equation
\begin{equation*}\label{INAR_model}
X_t = \alpha \odot  X_{t-1} + \epsilon_t = \sum_{j=1}^{X_{t-1}}N_{j} + \epsilon_t, \quad t\in\mathbb{Z},
\end{equation*}
where $\alpha \odot$ is a thinning operator,  $\alpha \in (0, 1)$, $\{\epsilon_t\}_{t\in \mathbb{Z}}$ is an innovation
sequence of independent and identically distributed non-negative integer-valued
random variables, not depending on past values of $\{X_{t}\}_{t\in \mathbb{Z}}$, mean $\mu_\epsilon ( < \infty)$
and variance $\sigma^2_{\epsilon} ( < \infty)$. It is also assumed that the $N_j$ variables
that define $\alpha \odot  X_{t-1}$ are independent of the variables from which
other values of the series are calculated, and are such that $\textrm{E}(N_j) = \alpha$. Moreover, we assume that all $N_j$ variables
defining the thinning operations are independent of the innovation sequence $\{\epsilon_t\}_{t\in \mathbb{Z}}$.
The autocorrelation function of $\{X_{t}\}_{t\in \mathbb{Z}}$ is of the same form as in the case of the usual AR(1) processes.

The stationary marginal distribution of $\{X_{t}\}_{t\in \mathbb{Z}}$ can be determined from the equation
\begin{equation}\label{pgfx}
\varphi_{X}(s) = \varphi_{X}(\varphi_{N}(s)) \cdot \varphi_{\epsilon}(s),
\end{equation}
where $\varphi_{X}(s) := \textrm{E}(s^{X})$, $\varphi_{N}(s) := \textrm{E}(s^{N})$ and $\varphi_{\epsilon}(s) := \textrm{E}(s^{\epsilon})$ denote the pgf's of $\{X_{t}\}_{t\in \mathbb{Z}}$, $\{N_{j}\}_{j=1}^{X_{t-1}}$
and $\{\epsilon_t\}_{t\in \mathbb{Z}}$, respectively.
Equation (\ref{pgfx}) can be used to obtain the distribution of the innovations sequence if
the marginal distribution of the observable INAR(1) process is known. By deriving the pgf's in (\ref{pgfx}), and using the stationarity of $\{X_{t}\}_{t\in \mathbb{Z}}$, it is easily shown that the pgf
of $\{\epsilon_t\}_{t\in \mathbb{Z}}$ is given by
\begin{equation*}\label{pgfe}
\varphi_{\epsilon}(s) = \frac{\varphi_{X}(s)}{\varphi_{X}(\varphi_{N}(s))}.
\end{equation*}

Next, we present some methods to obtain the probability
mass function of the innovation sequence of an INAR(1) process when we can rewrite the pgf
of the innovation process as a quotient of two polynomial functions of enequal degrees.

\subsection{Method 1: Fractional decomposition of the innovation distribution}

As in \citet{feller08aa}, let us assume that the pgf of the innovations process is a rational function given by
\begin{equation*}\label{0.0}
\varphi_{\epsilon}(s)=\frac{U_p(s)}{V_q(s)},
\end{equation*}
where $U_p(s)$ and $V_q(s)$ are two polynomials of degrees $p$ and $q\ (p<q)$, respectively,
and that the equation $V_q(s)=0$  has $q$ distinct roots $s_1,s_2,\ldots, s_q$ with $|s| < \min\{s_1,s_2,\ldots,s_q\}$. Then,  we readily have that $V_q(s)=\prod_{i=1}^q(s-s_i)$.
%

The probability mass function (pmf) of the innovation process can then be expressed~\citep[see][p. 276]{feller08aa} as
\begin{equation}\label{inn}
\textrm{Pr}[\epsilon_t=m]=\sum_{i=1}^q\frac{\rho_i}{s_i^{m+1}},
\end{equation}
where $\rho_i=\frac{-U_p(s_i)}{\frac{dV_q(s)}{ds}\mid_{s=s_i}}$.  If $s_1$ is smaller in absolute value than all other roots, the above pmf can be  approximated~\citep[see][p. 277]{feller08aa} by  $\frac{\rho_1}{s_1^{m+1}}$ as $m\longrightarrow\infty$.
Assuming that $p=q+r,  \ r\ge 0$, and using some well-known  algebraic manipulations,  it is possible to rewrite the pgf of the innovation process~\citep[see][p. 277]{feller08aa} as
\begin{equation}\label{frac}
\varphi_{\epsilon}(s)=U_r(s)+\frac{U_k(s)}{V_q(s)}, \quad k<q.
\end{equation}
Then, the results above can be applied to the rational  function $U_k(s)/V_q(s)$.
\begin{Remark}
Note that taking $s_i= (1+\mu_i)/u_i$ in (\ref{inn}),  the innovation process is a mixture of geometric distributions with parameters $\mu_i/(1+\mu_i)$  expressed as
\begin{equation}\label{inn_1}
\textrm{Pr}[\epsilon_t=m]=\sum_{i=1}^q c_i \left(\frac{\mu_i}{1+\mu_i}\right)^m \left(\frac{1}{1+\mu_i}\right),
\end{equation}
where $c_i = \rho_i \, \mu_i,\  i = 1, \ldots, q$.
We can find many different INAR(1) models proposed in the literature whose innovation processes are given in (\ref{inn_1}). These models will be called  fractional integer-valued autoregressive
 process and denoted here by FINAR(1) processes.
\end{Remark}
\begin{Remark}
\citet{feller08aa} has mentioned  in p. 276 that it is a hard work to find the exact  mixture distribution in (\ref{inn_1})  and a simple approximation could be of practical interest in order to get  satisfactory solutions  for inferential problems. In fact, if we suppose that $\mu_1$ is smaller than all the others, then as $m$ increases, we see that
$\textrm{Pr}[\epsilon_t=m]$ can be  approximated by the geometric distribution with parameter $\mu_1/(1+\mu_1)$.
\end{Remark}
Given the marginal pmf of INAR(0,1) process, the next remark presents in detail the proof of a recursive formula to obtain the innovation pmf. This alternative recursive procedure could be of interest if the pmf of the thinning operator is available.

\begin{Remark} (Alternative approach) The innovation pmf can be written by the recursive formula
\[
\textrm{Pr}[\epsilon_t=l] = -\frac{1}{b_0}\sum \limits_{i=l-q}^{l-1} c_i b_{l-i}, \quad l=q+1, q+2, \ldots.
\]

\begin{proof}
Suppose
$$\varphi_X(s)=U_p(s) = \sum\limits_{i=0}^{p}a_i s^i \quad \mbox{and} \quad \varphi_{\alpha \odot X}(s)=V_p(s) = \sum\limits_{j=0}^{q}b_j s^j\quad \text{with}\quad  p<q,$$
where $a_i$'s and $b_i$'s are some coefficients such that $a_0+a_1+\cdots+a_p=1$ and $b_0+b_1+\cdots+b_q=1$. Then, we readily have
\begin{equation}\label{eqI}
\varphi_\epsilon(s) \cdot \sum\limits_{j=0}^{q}b_j s^j=  \sum\limits_{i=0}^{p}a_i s^i,  \quad \forall s.
\end{equation}
Now, let
\[
\varphi_\epsilon(s) = \sum\limits_{l=0}^{\infty}c_j s^l,
\]
where $c_l$'s are some coefficients such that $\sum\limits_{l=0}^{\infty} c_l =1$. Then, upon substituting this in \eqref{eqI}, we obtain
\begin{equation}\label{eqII}
\sum\limits_{l=0}^{\infty}c_j s^l \cdot  \sum\limits_{j=0}^{q}b_j s^j=  \sum\limits_{i=0}^{p}a_i s^i,  \quad \forall s.
\end{equation}

Upon comparing coefficients of $s^0, s^1, \ldots, s^p$ on both sides of \eqref{eqII}, we obtain the following equations:
\begin{eqnarray*}
c_0b_0 &=& a_0, \\
c_0b_1 + c_1b_0 &=& a_1,\\
c_0b_2 + c_1b_1 + c_2b_0 &=& a_2,\\
&\vdots& \\
c_0 b_p + c_1b_{p-1}+c_2b_{p-2}+\cdots+c_{p}b_0 &=& a_p,
\end{eqnarray*}
which can be expressed equivalently as
\[
\begin{bmatrix}
b_0&0&0&\cdots&0\\
b_1&b_0&0&\cdots&0\\
b_2&b_1&b_0&\cdots&0\\
\vdots&\vdots&\vdots&\ddots&\vdots\\
b_p&b_{p-1}&b_{p-2}&\cdots&b_0
\end{bmatrix}
\begin{bmatrix}
c_0\\
c_1\\
c_2\\
\vdots\\
c_p
\end{bmatrix} =
\begin{bmatrix}
a_0\\
a_1\\
a_2\\
\vdots\\
a_p
\end{bmatrix}.
\]
This gives
\[
\begin{bmatrix}
c_0\\
c_1\\
c_2\\
\vdots\\
c_p
\end{bmatrix} = B_p^{-1}\begin{bmatrix}
a_0\\
a_1\\
a_2\\
\vdots\\
a_p
\end{bmatrix},
\quad
\text{where}
\quad
B_p = \begin{bmatrix}
b_0&0&0&\cdots&0\\
b_1&b_0&0&\cdots&0\\
b_2&b_1&b_0&\cdots&0\\
\vdots&\vdots&\vdots&\ddots&\vdots\\
b_p&b_{p-1}&b_{p-2}&\cdots&b_0
\end{bmatrix},
\]
which is clearly invertible since $b_0 \neq 0 \, (b_0 = V_q(0) >0)$. Next, upon comparing coefficients of $s^{p+1}, s^{p+2}, \ldots, s^{q}$ on both sides of
\eqref{eqII}, we get the following equations:
\begin{eqnarray*}
c_0b_{p+1} + c_1b_p + c_2b_{p-1} + \cdots+ c_{p+1}b_0 &=& 0, \\
c_0b_{p+2} + c_1b_{p+1} + c_2b_{p} + \cdots+ c_{p+2}b_0 &=& 0, \\
&\vdots& \\
c_0b_{q} + c_1b_{q-1} + c_2b_{q-2} + \cdots+ c_{q}b_0 &=& 0. \\
\end{eqnarray*}
This system of equations can be readily rewritten as follows
\begin{eqnarray*}
c_{p+1}b_0 &=& -(c_0b_{p+1} + c_1b_p + c_2b_{p-1} + \cdots + c_p b_1) , \\
c_{p+1}b_{1} + c_{p+2}b_0 &=&-(c_0b_{p+2} + c_1b_{p+1} + c_2b_{p} + \cdots + c_p b_2) , \\
&\vdots& \\
c_{p+1}b_{q-p-1}+ \cdots+c_qb_0&=& -(c_0b_{q} + c_1b_{q-1} + c_2b_{q-2} + \cdots + c_p b_{q-p}) . \\
\end{eqnarray*}
which can be expressed equivalently as
\[
\begin{bmatrix}
b_0&0&0&\cdots&0\\
b_1&b_0&0&\cdots&0\\
b_2&b_1&b_0&\cdots&0\\
\vdots&\vdots&\vdots&\ddots&\vdots\\
b_{q-p-1}&b_{q-p-2}&b_{q-p-3}&\cdots&b_0
\end{bmatrix}
\begin{bmatrix}
c_{p+1}\\
c_{p+2}\\
c_{p+3}\\
\vdots\\
c_q
\end{bmatrix} =
\begin{bmatrix}
b_{p+1}& b_p & b_{p-1}&\cdots&b_1\\
b_{p+2}& b_{p+1} & b_{p}&\cdots&b_2\\
b_{p+3}& b_{p+2} & b_{p+1}&\cdots&b_3\\
\vdots& \vdots & \vdots&\ddots&\vdots\\
b_{q}& b_{q-1} & b_{q-2}&\cdots&b_{q-p}\\
\end{bmatrix}
\begin{bmatrix}
c_0\\
c_1\\
c_2\\
\vdots\\
c_p
\end{bmatrix}.
\]
This readily yields
\[
\begin{bmatrix}
c_{p+1}\\
c_{p+2}\\
c_{p+3}\\
\vdots\\
c_q
\end{bmatrix} = -B_{q-p-1}^{-1} B^{*} B_{p}^{-1}\begin{bmatrix}
a_0\\
a_1\\
a_2\\
\vdots\\
a_p
\end{bmatrix},
\quad
\text{where}
\quad
B^{*} = \begin{bmatrix}
b_{p+1}& b_p & b_{p-1}&\cdots&b_1\\
b_{p+2}& b_{p+1} & b_{p}&\cdots&b_2\\
b_{p+3}& b_{p+2} & b_{p+1}&\cdots&b_3\\
\vdots& \vdots & \vdots&\ddots&\vdots\\
b_{q}& b_{q-1} & b_{q-2}&\cdots&b_{q-p}\\
\end{bmatrix}.
\]

Next, upon comparing coefficients of $s^{q+1}, s^{q+2}, \ldots,. $ on both sides of \eqref{eqII}, we get the following equations
\begin{eqnarray*}
c_1b_q + c_2 b_{q-1} + c_3b_{q-2} + \cdots + c_{q+1}b_0 &=& 0, \quad (s^{q+1}) \\
c_2b_q + c_3 b_{q-1} + c_4b_{q-2} + \cdots + c_{q+2}b_0 &=& 0, \quad (s^{q+2}) \\
c_3b_q + c_4 b_{q-1} + c_5b_{q-2} + \cdots + c_{q+3}b_0 &=& 0, \quad (s^{q+3}) \\
&\vdots&
\end{eqnarray*}
which readily yields the following solutions:
\begin{eqnarray*}
c_{q+1} &=& -\frac{1}{b_0} \sum\limits_{i=1}^{q}c_i b_{q-i+1},\\
c_{q+2} &=& -\frac{1}{b_0} \sum\limits_{i=2}^{q+1}c_i b_{q-i+2},\\
c_{q+3} &=& -\frac{1}{b_0} \sum\limits_{i=3}^{q+2}c_i b_{q-i+3},\\
&\vdots&
\end{eqnarray*}
and in general, we have
\[
c_{l} = -\frac{1}{b_0}\sum \limits_{i=l-q}^{l-1} c_i b_{l-i}, \quad l=q+1, q+2, \ldots
\]
\end{proof}
\end{Remark}

\begin{example}[Geometric INAR(1) process]
As in \citet{mckenzie85}, let us consider the marginal distribution of $\{X_{t}\}_{t\in \mathbb{Z}}$  to be a geometric distribution with parameter $\theta \in (0, 1)$ with pmf
$$\textrm{Pr}[X_t=m]=(1-\theta)^m\theta, \quad  m=0,1,2,\ldots$$
The GINAR(1) process is based on the binomial thinning operator (Steutel and Van Harn 1979), $\alpha\circ X_{t-1}:=\sum_{j=1}^{X_{t-1}}Y_{j}$, where the so-called counting series $\{Y_{j}\}_{j\geq 1}$ is a sequence of independent and identically distributed Bernoulli random variables with $\textrm{Pr}(Y_j = 1) = 1 - \textrm{Pr}(Y_j = 0) = \alpha \in [0, 1)$, and satisfies the equation
$$X_t=\alpha\circ X_{t-1}+\epsilon_t, \quad t \in \mathbb{Z}.$$

Thus, the pgf of the innovation processe can be  written as
$$\varphi_{\epsilon}(s)=\frac{\varphi_{X_t}(s)}{\varphi_{\alpha\circ X_{t-1}}(s)}=\frac{U(s)}{V(s)}, \quad |s| < 1/(1-\theta),$$
where
$$
U(s)= \theta+(1-\theta)\alpha-(1-\theta)\alpha\, s \quad \textrm{and} \quad
V(s)=1-(1-\theta)s.$$

Since we have here two linear functions with $p=q=1$, upon using a simple algebraic manipulation  suggested in (\ref{frac}), the pgf of the innovation process can be rewritten as
\begin{equation*}\label{ex_1.1}
\varphi_{\epsilon}(s)=\frac{U(s)}{V(s)}=\alpha+(1-\alpha)\frac{\theta}{1-(1-\theta)s}.
\end{equation*}

In this example, we have shown that the GINAR(1) process is a FINAR(1) process with the polynomial functions $U(s)$ and $V(s)$ having the same degrees $p=q=1$.  By using the rational decomposition in (\ref{inn})  for the fractional function $\frac{\theta(1-\alpha)}{1-(1-\theta)s}$, it can be seen that the innovation process follows a zero-inflated geometric distribution given by
$$\textrm{Pr}[\epsilon_t=m]=\alpha \mathbf{1}_{\{0\}}(m)+(1-\alpha)\theta(1-\theta)^m, \quad  m=0,1,2,\ldots$$
\end{example}

\begin{example}[New geometric INAR(1) process]
In this example, we consider the stationary NGINAR(1) process with  negative binomial thinning operator, introduced by \citet{Ristic:2009aa}, given by
$$X_t=\alpha * X_{t-1}+\epsilon_t, \quad t \in \mathbb{Z}, $$
where $\alpha * X_{t-1}=\sum_{i=1}^{X_{t-1}}W_i,\  \alpha \in [0,1)$, $\{W_i\}_{t \in \mathbb{Z}} $ is a sequence of iid random variables with geometric distribution, i.e., with pmf given by $\textrm{Pr}[W_i=m]=\left(\frac{\alpha}{1+\alpha}\right)^m\frac{1}{1+\alpha}$, and $\{X_{t}\}_{t\in \mathbb{Z}}$ is a stationary process with geometric$(\mu/(1 + \mu))$ marginals with pmf given by $\textrm{Pr}[X_t=m]=(\frac{\mu}{1+\mu})^m\frac{1}{1+\mu}, \mu > 0$. Using these assumptions of the NGINAR(1) process, the pgf of the innovation  sequence is given by
\begin{equation*}\label{ex1.2}
\varphi_{\epsilon}(s)=\frac{U(s)}{V(s)}, \quad |s| < 1,
\end{equation*}
where $U(s)=1+\alpha(1+\mu)-\alpha(1+\mu)s$ and $V(s)=\alpha\,\mu\left(s-\frac{1+u}{\mu}\right)\left(s-\frac{1+\alpha}{\alpha}\right)$.

Since,  NGINAR(1) process is a fractional INAR(1) process with $p=1$ and $q=2$, the goal is to apply the fraction decomposition  described  before to obtain the pmf of the innovation processe $\{\epsilon_t\}_{t \in \mathbb{Z}}$. The decomposition proceeds as follows:
\begin{itemize}
\item[i)] The roots of the quadratic function $V(s)$ are
$$s_1=\frac{1+\mu}{\mu} \quad  \text{and}\quad  s_2=\frac{1+\alpha}{\alpha};$$
\item[ii)] The coefficients $\rho_1$ and $\rho_2$ are
$$\rho_1=-\frac{U(s_1)}{V^{\prime}(s_1)}=\frac{1}{\mu}\left(1-\frac{\alpha\mu}{\mu-\alpha}\right) \quad  \text{and}\quad  \rho_2=-\frac{U(s_2)}{V^{\prime}(s_2)}=\frac{\mu}{\mu-\alpha}.$$
\end{itemize}
 Thus, the pmf of the innovation sequence is obtained from (\ref{inn}) and is given by
 \begin{equation*}\label{ex1.2.2}
\textrm{Pr}[\epsilon_t=m]=\left(1-\frac{\alpha\mu}{\mu-\alpha}\right)\left(\frac{\mu}{1+\mu}\right)^{m}\frac{1}{1+\mu}+\frac{\alpha\mu}{\mu-\alpha}\left(\frac{\alpha}{1+\alpha}\right)^{m}\frac{1}{1+\alpha},\quad   m=0,1,2,\ldots
 \end{equation*}
The pmf of the innovation sequence is well defined with the condition $\alpha \in [0, \mu/(1+\mu)]$ for $\mu > 0$ being necessary to guarantee that all probabilities are non-negative, and
is undefined for values outside  this range. Note that this result was obtained earlier in \citet{Ristic:2009aa}. .\\\\
\end{example}

\begin{example}[Dependent counting INAR(1) process]
\citet{Ristic:2013aa} proposed a geometrically distributed time series generated by dependent Bernoulli count series, called DCINAR(1) process.
The DCINAR(1) process is based on new generalized binomial thinning operator $\alpha \circ_\theta$
and satisfies the following equation:
\[
X_t=\alpha\circ_\theta X_{t-1}+\epsilon_t, \quad t \in \mathbb{Z},
\]
where the operator $\alpha \circ_\theta$ is defined as $\alpha\circ_\theta X = \sum_{i=1}^{X_{t-1}} U_i $, $\alpha, \theta \in [0,1]$, and $\{U_i\}_{i\in\mathbb{N}}$ is a sequence of Bernoulli($\alpha$) variables  with $\textrm{E}[U_i] = \alpha$ and $\textrm{Var}[U_i] = \alpha(1-\alpha)$. Moreover, these random variables are dependent, since $\textrm{Corr}[U_i,U_j]=\theta^2, \, \forall \, \theta\neq 0$ and $i\neq j$. For more details, see \citet{Ristic:2013aa}.

The pgf of the innovation sequence can be expressed into partial fractions as
\begin{equation}\label{ex3}
\varphi_{\epsilon}(s)=\frac{\alpha(1-\theta)(\alpha+\theta-\alpha\,\theta)}{\alpha+\theta-2\,\alpha\,\theta}+
\frac{\rho_1}{s_1-s}+\frac{\rho_2}{s_2-s}, \quad |s| < 1,
\end{equation}
where
$$
\rho_1=\frac{A_1}{\mu},\  \rho_2=\frac{A_2}{(\alpha+\theta-2\,\alpha\,\theta)\mu} \quad \textrm{and} \quad
s_1=\frac{1+\mu}{\mu},\ s_2=\frac{1+(\alpha+\theta-2\,\alpha\,\theta)\mu}{\alpha+\theta-2\,\alpha\,\theta)\mu}.
$$
The expressions for $A_1$ and $A_2$ are given in  \citet{Ristic:2013aa}. Taking $\mu_1=\mu$ and  $\mu_2=(\alpha+\theta-2\,\alpha\,\theta)\mu$,
the mixture innovation distribution in  \citet{Ristic:2013aa} follows from (\ref{inn_1}) and (\ref{ex3}) as
\begin{equation}\label{ex4}
\textrm{Pr}[\epsilon_t=m] = A_0\,\mathbf{1}_{\{0\}}(m)+A_1\left(\frac{\mu_1}{1+\mu_1}\right)^{m}\frac{1}{1+\mu_1}+A_2\left(\frac{\mu_2}{1+\mu_2}\right)^{m}\frac{1}{1+\mu_2},
\end{equation}
where
$$A_0=1-(A_1+A_2)=\frac{\alpha(1-\theta)(\alpha+\theta-\alpha\,\theta)}{\alpha+\theta-2\,\alpha\,\theta}, A_0 + A_1 + A_2 = 1 \quad \textrm{and} \quad A_i \geq 0, \, i = 0, 1, 2.$$
We note that the probability distribution in (\ref{ex4}) is a zero-inflated model involving two geometric distributions. This result is not cleary stated in Theorem 2 of \citet{Ristic:2013aa}.
\end{example}

\subsection{Method 2: Linear rational probability generation function}

In this section, we extend the innovation sequence of the geometric INAR(1) process discussed in Example 1. We consider here the a linear fractional  probability generating function with four-parameters given by
\begin{equation}\label{ex5}
\varphi_{\epsilon}(s)=\frac{a+b\,s}{c+d\,s}, \quad s \in [0,1] \quad \text{and} \quad \varphi_{\epsilon}(0) < 1,
\end{equation}
with the following parametric restrictions: $a<c$ and $a+b=c+d$.
After a simple algebraic manipulation, the pgf in (\ref{ex5}) can be decomposed into partial fractions as
\begin{eqnarray*}\label{ex6}
\varphi_{\epsilon}(s)&=&\frac{b}{d}+\frac{\rho}{s_1-s},
\end{eqnarray*}
where $s_1 = -c/d$ and $\rho= b\,c/d^2 - a/d$. Following the same procedure as in \citet[][p. 276]{feller08aa}, we obtain an exact expression for the pmf of the innovation sequence as
\begin{equation*}\label{ex7}
\textrm{Pr}[\epsilon_t=m]=\frac{b}{d} \mathbf{1}_{\{0\}}(m)+\frac{\rho}{s_1^{m+1}}, \quad m=0,1,2,\ldots
\end{equation*}
Taking $s_1 = 1/(1-\theta),\  0 \leq \theta < 1$, we have the pmf of the three-parameter innovation sequence
\begin{equation}\label{ex8}
\textrm{Pr}[\epsilon_t=m]=\frac{b}{d}\mathbf{1}_{\{0\}}(m)+\left(1-\frac{b}{d}\right)(1-\theta)^m\theta, \quad m=0,1,2,\ldots
\end{equation}
Note that the two-parameter innovation distribution in Example 1 is deduced by taking $b=-(1-\theta)\alpha$ and $d=-(1-\theta)$.
The mean and the variance of the innovation process in (\ref{ex8}) are given by
\begin{eqnarray*}
\textrm{E}[\epsilon_t]&=&\left(\frac{1-\theta}{\theta}\right)\left(1-\frac{b}{d}\right),\\
\textrm{Var}[\epsilon_t]&=&\left(\frac{1-\theta}{\theta}\right)\left(1-\frac{b}{d}\right)\left[
\frac{1}{\theta}\left(1+\frac{b}{d}\right) - \frac{b}{d}
\right],
\end{eqnarray*}
respectively.

The Fisher index of the dispersion of innovation process is given by
$$\mathbf{I}_{\epsilon}=\frac{\textrm{Var}[\epsilon_t]}{\textrm{E}[\epsilon_t]}= \frac{1}{\theta}\left(1+\frac{b}{d}\right) - \frac{b}{d}. $$
This Fisher index indicates equidispersion of the innovation process if $b = -d$, underdispersion if $b>-d$ and overdispersion if $b < -d$. Note that for $b = -d$, we have the mean to equal the variance, but the pmf in (\ref{ex8}) is not
the classical Poisson distribution.\\

\begin{example}[Two-parameter innovation sequence]
Let us now consider the following two-parameter pgf discussed by \cite{bouzar:2017}
\begin{equation*} \label{ex9}
\varphi_{\epsilon}(s) = 1-m\frac{1-s}{1+r(1-s)}, \quad  0\leq s\leq 1,
\end{equation*}
where $r\geq 0$ and $m\leq r+1$. It can be easily seen that the above pgf is a fraction linear function with
$a=1-r-m,\ b=m-r,\ c=1+r$ and $d=-r$. For this particular case we have $s_1 = (r+1)/r$ and $b/d = 1 - m/r$. The corresponding two-parameter innovation process is given by
\begin{equation*}\label{ex10}
\textrm{Pr}[\epsilon_t=m]=\left(1-\frac{m}{r}\right)\mathbf{1}_{\{0\}}(m)+\left(\frac{m}{r}\right)\left(\frac{1}{1+r}\right)\left(\frac{r}{1+r}\right)^m, \quad m=0,1,2,\ldots
\end{equation*}
\end{example}

\begin{example}[Zero-modified geometric innovation process]
\citet{Johnson2005} have formulated in Section 8.2.3 the following pgf
\begin{equation}\label{ex11}
\varphi_{\epsilon}(s) = 1 - k + k \frac{1}{1+\mu(1-s)}, \quad |s| < 1,
\end{equation}
with the parametric restriction $-1/\mu\leq k\leq 1$.
It is straightforward to see that the pgf in (\ref{ex11}) has a linear representation given by
$$
a = 1 + k\mu, \quad b =-k\mu, \quad c = 1 + \mu \quad \mbox{and} \quad  d = -\mu.
$$
Note that $s_1 = -c/d = (1+\mu)/\mu$ and the parametric restriction $0\leq a\leq c$ implies that $-1/\mu\leq k\leq 1$. Also, we have  $b/d=k$ in view of (\ref{ex8}), with $\theta = 1/(1+\mu)$, that the pmf of $\{\epsilon_t\}_{t \in \mathbb{Z}}$ is given by
$$\textrm{Pr}[\epsilon_t=m]=k\, \mathbf{1}_{\{0\}}(m)+(1-k)\left(\frac{1}{1+\mu}\right)\left(\frac{\mu}{1+\mu}\right)^m \quad m=0,1,\ldots$$
This pmf is the so-called zero-modified geometric distribution which presents equidispersion when $k = -1$; underdispersion when $\mu \in (0, 1)$ and $k \in [-1/\mu, -1)$; and overdispersion when $k \in (-1, 1)$. For $0 < k < 1$, it is known as zero-inflated distribution and zero-deflated distribution for $-1/\mu < k < 0.$ The zero-modified geometric distribution was considered in a recent paper by \cite{Bourguignon:2018}
to formulate a new INAR(1) process with zero-modified geometric innovations. Also, if we take $k=-\pi/\mu,\ 0 \leq \pi \leq 1$, the zero-modified geometric distribution is the Bernoulli-Geometric distribution with parameters $\mu$ and $\pi$ introduced by \citet{Bourguignon:2017ab}. This can be easily seen by taking the linear representation~\citep[see][]{Bourguignon:2017ab}
\begin{eqnarray*}\label{ex12}
a = 1 - \pi, \quad b = \pi, \quad c = 1 + \mu \quad \mbox{and} \quad d=-\mu.
\end{eqnarray*}
\end{example}

\subsection{Method 3: Quadratic rational probability generation function}

In this section, we consider the pgf of the innovation process as a quadratic rational function expressed as follows:
\begin{equation}\label{ex13}
\varphi_{\epsilon}(s)=\frac{a\,s^2 + b\,s + c}{a\,s^2 + \bar{b}\,s + \bar{c}}, \quad s \in [0,1] \quad \text{and} \quad \varphi_{\epsilon}(0) < 1,
\end{equation}
where $a\neq 0,\ b+c=\bar{b}+\bar{c}$ and $c\leq \bar{c}$.

Initially, to facilitate the partial fraction decomposition, we have to reduce the degree of the polynomial function in the numerator of (\ref{ex13}) by using a simple algebraic manipulation. Thus, we obtain a new expression for the pgf as
\begin{eqnarray}\label{ex14}
\varphi_{\epsilon}(s)&=&1+\frac{(b-\bar{b})s+c-\bar{c}}{as^2+\bar{b}s+\bar{c}}
=1+\frac{U_1(s)}{(s-s_1)(s-s_2)},
\end{eqnarray}
with
$$U_1(s)=\frac{(b-\bar{b})(s-1)}{a},$$
where $s_1$ and $s_2$ are two different roots of the polynomial function of the denominator in (\ref{ex13}) such that
$$a\,s^2+\bar{b}\,s+\bar{c}=a(s-s_1)(s-s_2).$$
In order to obtain the innovation distribution, from here on, we assume  that  $s_2\geq s_1>1$.
The rational function $\frac{U_1(s)}{(s-s_1)(s-s_2)}$ in (\ref{ex14}) can now be  decomposed into partial fractions  \cite[see][]{feller08aa} and the pgf of the innovation variable can be rewritten as
\begin{eqnarray*}\label{ex15}
\varphi_{\epsilon}(s)&=&1+\frac{\rho_1}{s_1-s}+\frac{\rho_2}{s_2-s},
\end{eqnarray*}
where
$$\rho_1 =-\frac{U_1(s_1)}{s_1-s_2} \quad \text{and} \quad \rho_2 =-\frac{U_1(s_2)}{s_2-s_1}.$$

Using the same geometric expansion for $\frac{1}{s_1-s}$ and $\frac{1}{s_2-s} $ as in \citet{feller08aa}, the pmf of the innovation sequence is obtained as
\begin{equation}\label{ex16}
\textrm{Pr}[\epsilon_t=m]=
\left\{\begin{array}{lcc}
\pi, &\textrm{if} &m=0,\\
(1-\pi)[w_1(1 - p_1)\,p_1^{m-1} + w_2(1 - p_2)\, p_2^{m-1}], &\textrm{if} & \ m\ge 1,
\end{array}\right.
\end{equation}
where $p_1 = 1/s_1$, $p_2 = 1/s_2 (p_2 \leq p_1)$ and $\pi = c/\bar c$. Note that $w_1 (1 - p_1)\,p_1^{m-1} + w_2 (1 - p_2)\, p_2^{m-1} $ is a mixture of two geometric distributions, where $w_1 = p_1/(p_1-p_2)$ and $w_2 = p_2/(p_2-p_1)$ are the weights with $w_1+w_2 =1$.
If $\pi \equiv p_2 \equiv 0$, then $\epsilon_t \sim \textrm{Geo}(1-p_1)$ and if $p_2 \equiv 0$, then $\epsilon_t$ has a hurdle geometric distribution with parameter $1 - p_1$. Also, if $\pi \equiv p_1 \equiv 0$, then $\epsilon_t \sim \textrm{Geo}(1-p_2)$ and if $p_1 \equiv 0$, $\epsilon_t$ has a hurdle geometric distribution with parameter $1 - p_2$.

Since in (\ref{ex16}) the zeros come from the Bernoulli  process with parameter $\pi$ and the nonzeros come from a different process    characterized  by a mixture of two geometric distributions, it  may be referred to as  ``hurdle-fractional geometric innovation distribution" and denoted by HFG.
The mean and the variance of the innovation process in (\ref{ex16}) are given by
\begin{eqnarray*}
\textrm{E}[\epsilon_t]&=& \frac{(1-\pi)}{(1-p_1)(1-p_2)},\\
\textrm{Var}[\epsilon_t]&=& \frac{(1-\pi) \left[p_1^2p_2-3p_1p_2+ p_1+ p_2+\pi \right]}{(1-p_1)^2 (1-p_2)^2},
\end{eqnarray*}
respectively.
%
%
\vspace{0.5cm}

\begin{example}[Inflated-parameter geometric INAR(1) process based on binomial thinning operator]

In this example, we present a new stationary first-order non-negative integer valued
autoregressive process, $\{X_{t}\}_{t\in \mathbb{Z}}$, with inflated-parameter geometric \citep{kolev2000} marginals.
The proposed process is based on the binomial thinning operator and satisfies the  equation:
\[
X_t = \alpha \circ X_{t-1} + \epsilon_t, \quad t \in \mathbb{Z},
\]
with $\alpha \in [0, 1)$ and $\{X_{t}\}_{t\in \mathbb{Z}}$ being a stationary process with inflated-parameter geometric marginals, i.e., with pmf given by
\begin{equation*}\label{ex17}
\textrm{Pr}[X_t=m]=\frac{\rho}{\mu+\rho} \mathbf{1}_{\{0\}}(m)+\left(1-\frac{\rho}{\mu+\rho}\right)\left(\frac{\mu+\rho}{1+\mu}\right)^m\left(\frac{1-\rho}{1+\mu}\right),\quad m=0,1,2,\ldots,
\end{equation*}
where $\rho \in [0, 1)$ and $\mu > 0$. The inflated-parameter geometric distribution is well-known as $\rho$-geometric distribution \citep[see][]{kolev2000} and  considered in a recent paper by  \citet{Borges:2017aa} to formulate a new thinning operator. The corresponding pgf is given by
\begin{equation*}\label{ex18}
\varphi_{X}(s)=\frac{1-\rho s}{1-s\rho+\mu(1-s)}, \quad |s| <1.
\end{equation*}

The mean and variance of the process $\{X_{t}\}_{t\in \mathbb{Z}}$ are given, respectively, by
$$\textrm{E}[X_t] = \frac{\mu}{1 - \rho} \quad \textrm{and} \quad \textrm{Var}[X_t] = \frac{\mu(1 + \mu + \rho)}{(1 - \rho)^2},$$

and so the dispersion index, denoted by $I_X$, becomes
$$I_X = \frac{1 + \alpha + \rho}{1-\rho} = 1 + \frac{2\,\rho + \mu}{1 - \rho} \geq 1 + \mu > 1,$$
that is, the proposed inflated-parameter geometric INAR(1) process accommodates overdispersion.
If $\rho = 0$, then the process $\{X_{t}\}_{t\in \mathbb{Z}}$ reduces to the NGINAR(1) process introduced by
\citet{Ristic:2009aa}; see Example 2. The additional parameter $\rho$ has a natural interpretation in terms of both the ``zero-inflated" proportion and
the correlation coefficient. For more details about the inflated-parameter geometric distribution, see \cite{kolev2000}.

The pgf of the innovation sequence is given by
\begin{equation}\label{ex19}
\varphi_{\epsilon}(s) = \frac{\varphi_{X}(s)}{\varphi_{\alpha \circ X}(s)}= \frac{(1-\rho \, s)\{1-\rho[1-\alpha(1-s)]+\alpha\,\mu(1-s)\}}{[1-s\,\rho+\mu(1-s)]\{1-\rho[1-\alpha(1-s)]\}}, \quad |s| < 1.
\end{equation}

The main goal here is to obtain the innovation sequence by applying the fractional decomposition method for the quadratic rational function in (\ref{ex19}).  After a simple algebraic manipulation, the pgf in (\ref{ex19}) can be rewritten as a quotient of two quadratic polynomial functions, where
$$a=\alpha\,\rho(\mu+\rho),  b = -[\rho(1-\rho)+\alpha(\mu+\rho)(1+\rho)],  c=1-\rho+\alpha(\mu+\rho),$$
$$\bar{b}=-\{(\mu+\rho)[1-\rho(1-\alpha)]+\rho\,\alpha(1+\mu)\} \quad \textrm{and} \quad  \bar{c}=(1+\mu)[1-\rho(1-\alpha)],$$
with $1-\rho+\alpha(\mu+\rho) = c \leq \bar c = (1+\mu)[1-\rho(1-\alpha)]$ and $\pi  = \frac{1-\rho+\alpha(\mu+\rho)}{(1+\mu)[1-\rho(1-\alpha)]}$. The roots of the quadratic  function in the denominator of the pgf in (\ref{ex19}) are given by
\begin{equation*}\label{ex21}
s_1=\frac{1+\mu}{\rho+ \mu} \quad \mbox{and} \quad  s_2=\frac{1-\rho(1-\alpha)}{\rho\alpha},
\end{equation*}
with $s_2\geq s_1> 1$, and consequently
\[
p_1 = \frac{\rho+\mu}{1+\mu}, \quad p_2 = \frac{\alpha \,\rho}{1-\rho(1-\alpha)}, \quad w_1 = \frac{(\rho+\mu)[1-\rho(1-\alpha)]}{(\rho+\mu)[1-\rho(1-\alpha)] - \alpha\rho(1+\mu)}   \]
and
\[w_2 = -\frac{\alpha \,\rho(1+\mu)}{(\rho+\mu)[1-\rho(1-\alpha)] - \alpha\,\rho(1+\mu)}.\]

The pmf of the innovation sequence is then obtained from (\ref{ex16}) and is given by
\begin{equation*}
\textrm{Pr}[\epsilon_t=m]=
\left\{\begin{array}{lcc}
\pi, &\textrm{if} &m=0;\\
(1-\pi)[w_1(1 - p_1)\,p_1^{m-1} + w_2(1 - p_2)\, p_2^{m-1}], &\textrm{if} & \ m\ge 1.
\end{array}\right.
\end{equation*}

The mean and variance of $\{\epsilon_{t}\}_{t\in\mathbb{Z}}$ are given by

\[\textrm{E}[\epsilon_t] = \frac{\mu(1-\alpha)}{1-\rho} \]
and
\begin{gather*}
\textrm{Var}[\epsilon_t] =\frac{\mu(1-\alpha)\left[\mu ^2 (\alpha -\rho +1)+\mu  \left(\alpha  \rho +\alpha -\rho ^2-\rho +2\right)+\alpha  \rho  \left(\rho ^2-2 \rho +2\right)-\rho ^2+1\right]}{(1+\mu)(1-\rho)^3}.
\end{gather*}

\end{example}

\begin{example}[Hurdle geometric INAR(1) process based on binomial thinning operator]

In this example, we propose a new stationary first-order non-negative integer valued
autoregressive process, $\{X_{t}\}_{t\in \mathbb{Z}}$, with hurdle geometric marginals.
The proposed process is based on the binomial thinning operator and satisfies the equation
\begin{equation}\label{ex7}
X_t = \alpha \circ X_{t-1} + \epsilon_t, \quad t \in \mathbb{Z},
\end{equation}
with $\alpha \in [0, 1)$ and $\{X_{t}\}_{t\in \mathbb{Z}}$ being a stationary process with hurdle geometric marginals, i.e., with pmf given by
\begin{equation}
\label{bernoulli}
\textrm{Pr}(X_t=m)=\left\{
\begin{array}{lcl}
1-\mu&,& m=0, \\
\mu\left(\frac{\rho}{1+\rho}\right)^{m-1}\left(\frac{1}{1+\rho}\right) &,& m=1,2,\ldots, \\
\end{array}
\right.
\end{equation}
where $\mu, \rho \in (0,1)$ and its corresponding pgf is given by
\begin{eqnarray*}\label{ex24}
\varphi_{X_t}(s) = \frac{1 - (1-s)(\mu+\mu\,\rho - \rho)}{1+\rho-\rho\, s}.
\end{eqnarray*}
Consequently, the pgf of $\alpha\circ X_{t-1}$ is given by
\begin{eqnarray*}\label{ex25}
\varphi_{\alpha \circ X_t}(s) = \varphi_{X}(\varphi_{Y}(s)) = \frac{1 - \alpha(1-s)(\mu+\mu\,\rho - \rho)}{1+\alpha \rho-\alpha\,\rho \,s},
\end{eqnarray*}
where $\textrm{Pr}(Y = 1) = 1 - \textrm{Pr}(Y = 0) = \alpha \in [0, 1)$.

The hurdle geometric distribution is well-known as inflated-parameter Bernoulli distribution and $\rho$-Bernoulli distribution \citep[see][]{kolev2000} and has also been discussed in a recent paper by \citet{Borges:2016aa} to formulate a new thinning operator. If $\rho = 0$ in (\ref{bernoulli}), the hurdle geometric distribution coincides with the usual Bernoulli distribution with parameter $\mu$, taking values 0 and l, while if we put $\mu = \rho/(1 + \rho)$, the pmf of the geometric distribution with mean $\rho$ is obtained.

The mean and variance of $\{X_{t}\}_{t\in \mathbb{Z}}$ are given, respectively, by
\begin{equation*}\label{meanvarY}
\textrm{E}[X_{t}] = \alpha(1 + \rho) \quad \textrm{and} \quad \textrm{Var}[X_{t}]  = \mu(1+\rho)[\rho + (1 + \rho)(1 - \mu)],
\end{equation*}

and so the dispersion index is given by
$$I_X = \rho + (1 + \rho)(1 - \mu),$$
that is, the hurdle geometric distribution has equidispersion/overdispersion/underdispersion according to the following conditions (for fixed $0 < \mu < 1$):
\begin{eqnarray*}
  \rho = \frac{\mu}{2 - \mu} &\Rightarrow& \textrm{equidispersion},\\
  \frac{\mu}{2 - \mu} < \rho < 1 &\Rightarrow& \textrm{overdispersion},\\
 0 \leq \rho < \frac{\mu}{2 - \mu} &\Rightarrow& \textrm{underdispersion}.
\end{eqnarray*}

Now, let us derive the distribution of the random variable $\{\epsilon_t\}_{t\in \mathbb{Z}}$. Let $\varphi_{X}(s)$, $\varphi_{Y}(s)$ and $\varphi_{\epsilon}(s)$ be the pgf's of the random variables $X_{t}$, $Y_j$ and $\epsilon_{t}$, respectively. Since the process $\{X_{t}\}_{t\in\cal{Z}}$ is a stationary process, it follows from (\ref{ex7}) that
\begin{eqnarray}\label{ex26}
\varphi_{\epsilon}(s) =   \frac{\varphi_{X_t}(s)}{\varphi_{\alpha\circ X_{t-1}}(s)}= \frac{1 - (1-s)(\mu+\mu\,\rho - \rho)}{1+\rho-\rho s} \cdot  \frac{1+\alpha \rho-\alpha\,\rho\, s}{1 - \alpha(1-s)(\mu+\mu\,\rho - \rho)}, \quad |s|<1.
\end{eqnarray}

After a simple algebra, we can identify the polynomial functions with  corresponding coefficients
\begin{eqnarray*}
a &=& \alpha \,\rho [\rho - \mu(1 +  \rho)], \mbox{with} \; a>0 \; \mbox{if} \; \mu < \rho/(1+\rho),\\
b &=& - \left\{ 2 \alpha\, \rho [\rho - \mu(1 +  \rho)] + [\rho - \mu(1 +  \rho)]  + \alpha\,\rho \right\}, \\
c &=& 1 + \alpha\, \rho [\rho - \mu(1 +  \rho)] + [\rho - \mu(1 +  \rho)] + \alpha\, \rho = (1+\rho)( 1+ \alpha[\rho - \mu(1 +  \rho)] - (1-\alpha)\mu),  \\
\bar b &=& - \left\{  2 \alpha \,\rho [\rho - \mu(1 +  \rho)] + \alpha [\rho - \mu(1 +  \rho)] + \rho \right\},\\
\bar c &=& 1 + \alpha \,\rho [\rho - \mu(1 +  \rho)] + \alpha [\rho - \mu(1 +  \rho)] + \rho = (1+\rho)\{1+\alpha[\rho - \mu(1 +  \rho)]\}.
\end{eqnarray*}
 Note that $b+c = \bar b + \bar c = 1 - \alpha \rho [\rho - \mu(1 +  \rho)],\
c \leq \bar c $, and the roots of polynomial function in the denominator of (\ref{ex26}) are

\[
s_1 =\frac{1+\rho}{\rho} \quad \mbox{and} \quad s_2 =  \frac{1+\alpha[\rho-\mu(1+\rho)]}{\alpha[\rho-\mu(1+\rho)]},\]
with  $s_2\geq s_1> 1$. From (\ref{ex16}), the pmf of the innovation sequence is then obtained as
\begin{equation*}
\textrm{Pr}[\epsilon_t=m]=
\left\{\begin{array}{lcc}
\pi, &\textrm{if} &m=0;\\
(1-\pi)[w_1(1 - p_1)\,p_1^{m-1} + w_2(1 - p_2)\, p_2^{m-1}], &\textrm{if} & \ m\ge 1,
\end{array}\right.
\end{equation*}
with $\pi = 1 - (1-\alpha) \frac{\mu}{1+ \alpha[\rho - \mu(1 +  \rho)] }$, $p_1=\frac{\rho}{1+\rho}$, $p_2= \frac{\alpha[\rho-\mu(1+\rho)]}{1+\alpha[\rho-\mu(1+\rho)]}$, $w_1=\frac{\rho(1+\alpha[\rho-\mu(1+\rho)])}{\rho-\alpha[\rho-\mu(1+\rho)]}$
and $w_2 = - \frac{\alpha(1+\rho)[\rho-\mu(1+\rho)]}{\rho-\alpha[\rho-\mu(1+\rho)]}$.

The mean and variance of $\{\epsilon_{t}\}_{t\in\mathbb{Z}}$ are given by

$$\textrm{E}[\epsilon_t] = \mu(1-\alpha)(1+\rho),$$
and
$$\textstyle \textrm{Var}[\epsilon_t] =\mu(1-\alpha)\left[\alpha\rho^3 -\mu(1+\rho) \left[\alpha\left(\rho ^2+\rho +1\right)+\rho +1\right]  + 2(1+\alpha)\rho^2 + (2\alpha +3)\rho +1\right].$$

\end{example}

\subsection{Method 4: A flexible quadratic rational probability generation function}
In this subsection, we extend the Method 3 to include a new family of innovation distributions by allowing  the coefficients of the quadratic functions to be different. In this general case, the pgf will be of the form
\[
\varphi_{\epsilon}(s)=\frac{a\,s^2+b\,s+c}{\bar{a}\,s^2+\bar{b}\,s+\bar{c}}, \quad s \in [0,1] \quad \text{and}\quad \varphi_{\epsilon_t}(0)<1,
\]
with the restrictions $c\leq \bar{c}$ and $a+b+c=\bar{a}+\bar{b}+\bar{c}$.
Since the rational pgf above involves two quadratic polynomial functions we need to reduce the degree of the quadratic function of the numerator which, after a simple algebra ,can be rewritten as
\[\varphi_{\epsilon}(s)=\frac{a}{\bar{a}}+ \frac{U_1(s)}{V_2(s)},\]
where
\begin{equation*}\label{ex27}
U_1(s)=\frac{1}{\bar{a}}\left[\left(b-\frac{a\,\bar{b}}{\bar{a}}\right)s+\left(c-\frac{a\,\bar{c}}{\bar{a}}\right)\right] \quad \textrm{and} \quad
V_2(s)=\bar{a}\,s^2+\bar{b}\,s+\bar{c}=\bar{a}(s-s_1)(s-s_2).
\end{equation*}
Note that we are assuming that the quadratic function $V_2(s)$ has two distinct roots $s_1$ and $s_2$. Then, the rational function $\frac{U_1(s)}{V_2(s)}$ can be decomposed into partial fractions a
\begin{equation*}
\frac{U_1(s)}{V_2(s)}=\frac{\rho_1}{(s_1-s)}+\frac{\rho_2}{(s_2-s)},
\end{equation*}
where
$\rho_1=-\frac{U_1(s_1)}{s_1-s_2}\ \text{and} \ \rho_2=-\frac{U_1(s_2)}{s_2-s_1}.$

Using the same fractional expansion as the one the one in Method 3, the innovation distribution referred to as ``Modified hurdle-fractional geometric distribution" (MHFG) is given by


\begin{equation}\label{ex18}
\textrm{Pr}[\epsilon_t=m]=
\left\{\begin{array}{lcc}
\pi, &\textrm{if} &m=0,\\
(1-\pi)[w_1(1 - p_1)\,p_1^{m-1} + w_2(1 - p_2)\, p_2^{m-1}], &\textrm{if} & \ m \ge 1,
\end{array}\right.
\end{equation}
where
\[
p_1=\frac{1}{s_1},\quad  p_2=\frac{1}{s_2},\quad  w_1=\frac{\rho_1\, p_1^2}{(1-p_1)(1-\pi)} \quad \text{and} \quad w_2=\frac{\rho_2\, p_2^2}{(1-p_2)(1-\pi)}.
\]

The MHFG innovation distribution is a two-stage distribution where the first stage is characterized by a Bernoulli variable with parameter $\pi=c/\bar{c}$ and the second stage is a mixture of two geometric distributions. It is not difficult to see that if $a=\bar{a}$, the MHFG becomes the HFG obtained by Method 3.  The mean and the variance of the innovation process in (\ref{ex16}) are given by

\[\textrm{E}[\epsilon_t]= (1-\pi)\left\{ w_1 \cdot \frac{1}{(1-p_1)} + w_2 \cdot \frac{1}{(1-p_2)}\right\} = (1-\pi)\mu_{Z}\]
and

\begin{eqnarray*}
\textstyle
\textrm{Var}[\epsilon_t]&=&(1-\pi)\left\{ w_1\cdot \left[ \left(\frac{1}{1-p_1} - \mu_{Z}\right)^2 + \frac{p_1}{(1-p_1)^2}   \right] + w_2\cdot \left[ \left(\frac{1}{1-p_2} - \mu_{Z}\right)^2 + \frac{p_2}{(1-p_2)^2}   \right]\right\}\\
&& + \pi(1-\pi)\mu_{Z}^2,\\
&=& (1-\pi)\left\{\frac{w_1(1-w_1)(p_1-p_2)^2 + w_1(p_1-p_2) + p_2(1-p_1)^2}{(1-p_1)^2(1-p_2)^2}  + \pi \mu_{Z}^2\right\},
\end{eqnarray*}
respectively.

\begin{example}[Inflated-parameter geometric INAR(1) process based on NB thinning operator]

A discrete-time non-negative integer-valued stochastic process $\{X_{t}\}_{t\in\mathbb{Z}}$
is said to be an INAR(1) process with inflated-parameter geometric based on NB thinning operator if it satisfies the equation $(0 \leq \alpha < 1)$
\[
X_t = \alpha*X_{t-1} + \epsilon_t, \quad t \in \mathbb{Z},
\]
where `*' is the negative binomial thinning operator defined in Example 2.
We assume that the random variables of the process are identically distributed. If $\rho = 0$, then the process $\{X_{t}\}_{t\in \mathbb{Z}}$ is reduced to the NGINAR(1) process introduced by
\citet{Ristic:2009aa}; see Example 2. Thus, the pgf of the innovation sequence is given by
\begin{eqnarray}\label{eq19}
\varphi_{\epsilon}(s) = \frac{\varphi_{X_t}(s)}{\varphi_{\alpha * X}(s)} = \frac{1-s\,\rho}{1-s\,\rho+\mu(1-s)} \cdot \frac{1 + \alpha(1-s)-\rho+\alpha\,\mu(1-s)}{1+\alpha(1-s)-\rho}, \quad |s| < 1.
\end{eqnarray}

After a simple algebraic manipulation, the pgf in (\ref{eq19}) can be rewritten as a quotient of two quadratic polynomials  where $a = \alpha\, \rho (1+ \mu), \quad  b = -[\alpha(1+\mu)\rho +\alpha(1+\mu)+(1-\rho)\rho], \quad c = (1+\mu)[1-\rho + \alpha] - \mu(1-\rho), \quad
\bar a = \alpha (\rho + \mu), \quad
\bar b = -[\rho + \mu + \alpha + \alpha\,\rho + 2\,\alpha\,\mu - \mu\,\rho - \rho^2], \text{and}\quad
\bar c = (1+\mu)[1-\rho + \alpha].
$

Also, note that $a+b+c = \bar a + \bar b + \bar c = (1-\rho)^2$, $(1+\mu)[1-\rho + \alpha] - \mu(1-\rho) = c \leq \bar c = (1+\mu)[1-\rho + \alpha]$ and the roots of the polynomial function in the denominator of (\ref{eq19})  are
\[
s_1 = \frac{1+\mu}{\rho+\mu} \quad \mbox{and} \quad s_2 = \frac{1-\rho + \alpha}{\alpha}
\]
with $s_2 \geq s_1 > 1$. From (\ref{ex18}), the pmf of the innovation sequence is then given by
\begin{equation*}
\textrm{Pr}[\epsilon_t=m]=
\left\{\begin{array}{lcc}
\pi, &\textrm{if} &m=0,\\
(1-\pi)[w_1(1 - p_1)\,p_1^{m-1} + w_2(1 - p_2)\, p_2^{m-1}], &\textrm{if} & \ m\ge 1,
\end{array}\right.
\end{equation*}
where
\[
\pi = 1- \frac{\mu(1-\rho)}{(1+\mu)[1-\rho + \alpha]}, \quad p_1=\frac{\rho + \mu}{1+\mu},\quad  p_2=\frac{\alpha}{1-\rho + \alpha},\quad  w_1=\frac{(\alpha -\rho +1) (\alpha  \mu +\alpha -\mu -\rho )}{(\rho -1) (-\alpha +\mu +\rho )}
\]
and
\[\quad w_2=\frac{\alpha(1+\mu) (\alpha-\rho )}{(\rho -1) (\alpha -\mu -\rho )}.\]
\end{example}

The mean and variance of $\{\epsilon_{t}\}_{t\in\mathbb{Z}}$ are given by

$$\textrm{E}[\epsilon_t] = \frac{\mu(1-\alpha)}{1-\rho},$$
and
 $$\textrm{Var}[\epsilon_t] =\frac{\mu  \left[\alpha ^2 (1+\mu)^2 (\rho-2)+\alpha\left(\mu^2-\mu\left(\rho^2-4\rho+1\right)+2\rho-1\right) + (1+\mu)(1-\rho)(1+\mu +\rho)\right]}{(1+\mu) (1-\rho)^3} .$$

%
%

%
\begin{example}[Hurdle geometric INAR(1) process based on NB thinning operator]

Finally, in this example, we consider a new stationary INAR(1) process with hurdle geometric marginals based on NB thinning operator.
The proposed process satisfies the equation
\[
X_t = \alpha*X_{t-1} + \epsilon_t, \quad t\in\mathbb{Z},
\]
where $0 \leq \alpha < 1$. The pgf of the innovation sequence is given by
\begin{eqnarray}\label{pgf:ex9}
\varphi_{\epsilon}(s) = \frac{1- (1-s)[\mu(1+\rho) - \rho]}{1 + \rho(1-s)} \cdot \frac{1 + \alpha(1-s) + \rho\,\alpha(1-s)}{1 + \alpha(1-s)-\alpha(1-s)[\mu(1+\rho) - \rho]}.
\end{eqnarray}

Then, after a simple algebraic manipulation, the pgf in (\ref{pgf:ex9}) can be rewritten as a quotient of two quadratic polynomials  where:
$a=\alpha  (1+\rho )[(1+\rho) (1-\mu)-1]$,
$b=-[2 \, \alpha(1+\rho) [\rho -\mu(1+\rho)] + \alpha(1+\rho) - \mu(1+\rho)+\rho]$,
$c=(1+\rho)[\alpha(1+\rho)(1-\mu)+1-\mu]$,
$\bar a=\alpha\,\rho(1+\rho)(1-\mu)$,
$\bar b=-[\alpha(1+\rho)(1-\mu)(2\rho+1) + \rho]$ and
 $\bar c=(1+\rho)[\alpha(1+\rho)(1-\mu)+1]$.
Note that  $a+b+c = \bar a + \bar b + \bar c = 1$ and $c \leq \bar c$. Moreover, the roots of the polynomial function in the denominator of (\ref{pgf:ex9}) are
\[
s_1 = \frac{1+\rho}{\rho} \quad \mbox{and} \quad s_2 = \frac{1+\alpha(1+\rho)(1-\mu)}{\alpha(1+\rho)(1-\mu)}.
\]
with $s_2 \geq s_1 > 1$. From (\ref{ex18}), the pmf of the innovation sequence is then given by
\begin{equation*}
\textrm{Pr}[\epsilon_t=m]=
\left\{\begin{array}{lcc}
\pi, &\textrm{if} &m=0,\\
(1-\pi)[w_1(1 - p_1)\,p_1^{m-1} + w_2(1 - p_2)\, p_2^{m-1}], &\textrm{if} & \ m\ge 1,
\end{array}\right.
\end{equation*}
where
\[
\pi = \frac{\alpha  (1-\mu ) (\rho +1)-\mu +1}{\alpha(1-\mu ) (\rho +1)+1}, \quad p_1=\frac{\rho }{1+\rho},\quad  p_2=\frac{\alpha(1+\rho)(1-\mu)}{1+\alpha(1+\rho)(1-\mu)},\]
\[ w_1=\frac{(\alpha \, \rho +\alpha -\rho ) [\alpha  (\mu - 1) (\rho +1)-1]}{\alpha  (\mu -1) (\rho +1)+\rho } \quad \textrm{and} \quad w_2=-\frac{\alpha  (\rho +1) [\alpha  (\mu -1) (\rho +1)-\mu  (\rho +1)+\rho ]}{\alpha  (\mu -1) (\rho +1)+\rho }.\]
\end{example}

The mean and variance of $\{\epsilon_{t}\}_{t\in\mathbb{Z}}$ are given by

$$\textrm{E}[\epsilon_t] = \mu(1-\alpha)(1+\rho)$$
and
$$\textrm{Var}[\epsilon_t] =\mu(1+\rho) \left[\alpha ^2 (1+\rho) (\mu  \rho +\mu -\rho -2)+\alpha  \left[(1-\mu) \rho ^2-1\right]-\mu  (1+\rho)+2 \rho +1\right] .$$

\section{Concluding remarks}
In the recent literature concerning INAR(1) models, many papers have assumed a known marginal distribution but usually it is difficult to understand how the innovation distribution was obtained. In this paper, a new technique based on the fractional approach developed in
\citet[see][p. 276]{feller08aa}, is formulated to find the innovation process by using a simple algebraic manipulations. This fractional procedure has been discussed in detail for the linear and quadratic probability generating functions and illustrated with many recent INAR(1) models. We also see that some of these examples reproduce some new innovation processes which could be of interest in this area. For example, a new innovation process can be obtained if we assume an inflated-parameter geometric marginal distribution for the INAR(1) model as described in  Method 4. As part of our future research study,  we plan to study the four newly proposed processes in detail and their properties, the recursive method and associated inferential issues.


\end{document}